# Label-free biochemical quantitative phase imaging with mid-infrared photothermal effect

Miu Tamamitsu,[1] Keiichiro Toda,[1] Hiroyuki Shimada,[2] Takaaki Honda,[3] Masaharu Takarada,[3] Kohki Okabe,[3] Yu Nagashima,[4] Ryoichi Horisaki,[5,6] and Takuro Ideguchi[1,2,6,*]

[1]Department of Physics, The University of Tokyo, Tokyo 113-0033, Japan
[2]Institute for Photon Science and Technology, The University of Tokyo, Tokyo 113-0033, Japan
[3]Department of Pharmaceutical Sciences, The University of Tokyo, Tokyo 113-0033, Japan
[4]Department of Neurology, The University of Tokyo, Tokyo 113-0033, Japan
[5]Graduate School of Information Science and Technology, Osaka University, Osaka 565-0871, Japan
[6]PRESTPO, Japan Science and Technology Agency, Saitama 332-0012, Japan
[*]Corresponding author: ideguchi@ipst.s.u-tokyo.ac.jp

**Label-free optical imaging is valuable in biology and medicine with its non-destructive property and reduced optical and chemical damages. Quantitative phase (QPI) and molecular vibrational imaging (MVI) are the two most successful label-free methods, providing morphology and biochemistry, respectively, that have pioneered numerous applications along their independent technological maturity over the past few decades. However, the distinct label-free contrasts are inherently complementary and difficult to integrate due to the use of different light-matter interactions. Here, we present a unified imaging scheme that realizes simultaneous and in-situ acquisition of MV-fingerprint contrasts of single cells in the framework of QPI utilizing the mid-infrared photothermal effect. The fully label-free and robust integration of subcellular morphology and biochemistry would have important implications, especially for studying complex and fragile biological phenomena such as drug delivery, cellular diseases and stem cell development, where long-time observation of unperturbed cells are needed under low phototoxicity.**

Optical imaging is indispensable in biological and medical applications with its non-destructive property. Label-free imaging such as quantitative phase (QPI) and molecular vibrational imaging (MVI) is particularly valuable for studying fragile systems where exogenous labelling that often spoils the sample is not preferred[1-18]. QPI yields the sample-specific 2D optical-phase-delay[1] or 3D refractive-index (RI) distribution[2], which is the fundamental quantity used to visualize morphology of transparent samples as in the cases of the dark-field, phase-contrast and differential-interference-contrast microscopy. Compared to these traditional methods where the optical wavefront is perturbed to couple its phase into the amplitude, the direct quantification of the phase-delay or RI by QPI reveals the natural morphology that is more smooth and higher in spatial resolution and contrast, allowing for accurate and automatic cellular profiling[3]. Its essential capability is to translate the measured QP into the cellular dry-mass density, which has enabled quantification of cellular growth rate[4] and numerous other applications[5]. On the other hand, MVI yields the comprehensive spectroscopic information of biomolecular bonds based on Raman scattering[6-10] or mid-infrared (MIR) absorption[11-18]. Coherent Raman[7-10] and, more recently, MIR-photothermal[12-17] and -photoacoustic[18] imaging have gained attention due to the high spatial resolution and detection sensitivity. The state-of-the-art systems can perform video-rate imaging of, e.g., intracellular proteins, lipids or nucleic acids, allowing for high dimensional metabolic analysis[10], etc.

In spite of their independent technological maturity, the inherent limitations of MVI and QPI are still left unresolved. The RI does not offer chemical specificity[5] whereas the MV can be detected at limited regions where resonant molecules exist. It is also difficult to estimate the quantity of each biochemical constituent based on MV spectroscopy without the additional knowledge of the interaction lengths, the attenuation or scattering cross-sections[6], and the molecular masses of unknown biological compositions. Combining the two complementary information, i.e., quantitative morphology and qualitative biochemistry, can not only mitigate these limitations, but also synergistically expand the capability of label-free imaging. Indeed, multi-modal spontaneous Raman-QPI has shown potential to decompose the local cellular dry-mass density into the individual biochemical constituents[19]. However, merely combining independent modalities is not a robust solution because there are always mismatches in resolutions, sampling points, fields of view (FOVs), etc. For instance, in the spontaneous Raman-QPI system, the lack of depth-resolution in the QPI prohibits accurate decomposition of the dry-mass density into independent biomolecular components, whereas the slow acquisition speed of the spontaneous Raman imaging is accompanied by motion-blur artifacts of cellular dynamics. Accurate correlation of the spatiotemporal evolutions of subcellular morphology and biochemistry yields more comprehensive and robust pictures of complex biological systems, and a fully label-free implementation would have important implications when studying fragile phenomena.

Here, we present a unified imaging scheme that bridges this technological gap between the two label-free modalities, realizing simultaneous and in-situ acquisition of MV contrasts in the framework of QPI using the MIR photothermal effect. Preliminary results on this method, which we term MV-sensitive QPI (MV-QPI), have been recently reported[20]. The focus of this work is to further develop the MV-QPI method, proving its practical bioimaging capability in the broadband MIR fingerprint region while also pioneering the depth- and super-resolved imaging performance beyond the diffraction limit posed in other MVI techniques[6-18]. To highlight MV-QPI's versatility, we present two implementations of QPI scheme for the single-cell imaging application. The first is based on digital holography (DH) where we demonstrate live-cell 2D MV-QPI. The second is based on optical diffraction tomography (ODT) where we demonstrate, for the first time to our knowledge, the depth-resolved MV-QPI. We expect our MV-QPI method allows for merging of the independent knowledge of the QPI and MVI communities, leading to more comprehensive understanding of various biological phenomena.

**Results**
**MV-QPI.** The concept of MV-QPI is illustrated in Fig. 1. MV-QPI is a super-resolved MIR imaging method based on the wide-field visible (VIS) QPI detection of site-specific RI changes induced by the MIR photothermal effect. The entire volume of the sample placed in the objective focus of a QPI system is illuminated by MIR light of a certain wavenumber that excites the resonant molecular species to their fundamental vibrational states (see Fig. 1a). Through the non-radiative decay of the MVs, the local RIs in the vicinities of the resonant molecules decrease due to the rise of the temperature[12-17], which are detected by QPI with the spatial resolution of the VIS probe light[20]. The obtained MIR "OFF" QPI reveals the quantitative morphology of the sample, while the subtraction of the "OFF" from the "ON" QPI reveals the site-specific phase or RI decrease which reflect the local MIR absorption property (see Fig. 1b). Scanning the MIR wavenumber yields spectroscopic images of different MV resonances. The broadband MIR absorption spectrum can be obtained at each spatial point, which can be used to identify the local molecular compositions through chemometric analysis. Eventually, we can map the MV-based biomolecular distributions within the global morphology of the sample provided by the QP contrast.

Generally, the MIR photothermal imaging[12-17] including our MV-QPI method[20] offers (1) the high MV detection sensitivity based on the MIR absorption having ~8 orders of magnitude larger cross-section compared to Raman scattering[17] and (2) the low photodamage by the use of the low-photon-energy MIR excitation that most unlikely excites the electronic transitions[21] of biomolecules. In this work, we additionally harness the capability of QPI to computationally synthesize the 3D spatial-frequency aperture of the imaging system[2,22]. This enables us to further achieve (3) the higher lateral spatial resolution by the expanded bandwidth of the synthetic aperture that surpasses the numerical aperture (NA) of a single objective lens which poses the diffraction limit in other far-field MVI techniques[6-18,20] and (4) decoupling the undesired MIR absorption effect of the surrounding aqueous medium, which is the well-known problem of MIR microscopy in general, by the depth-resolving power offered by the axial bandwidth of the 3D synthetic aperture.

Figures 1c and 1d describe the mechanism of the diffraction limit when the sample is illuminated with a coherent wavefront[2,22,23]. Figure 1c illustrates the case of standard 2D QPI where the synthetic aperture technique is not used[20]. The object is illuminated with an orthogonal plane wave where a higher-frequency structure diffracts the wavefront with a larger angle. The objective lens collects the transmitted field with a limited angular range determined by its NA, posing the diffraction limit due to the low-pass filtering effect. Other MVI techniques that use focused optical probe[6-18] share the same diffraction-limit mechanism posed by the NA of a single objective lens. In the synthetic-aperture QPI, this limitation can be surpassed using tilted plane wave illumination[2,22,23] (see Fig. 1d). The propagation wavevector of the diffracted wavefront is accordingly tilted, such that the higher-frequency contents can be collected with the same objective lens. Essentially, the tilt shifts the location of the objective lens' NA in the spatial-frequency domain. The Fourier diffraction theorem[24] also allows us to map the 2D frequency aperture to the 3D frequency space when the illumination is monochromatic, which becomes a spherical cap called the Ewald's sphere as shown in Figs. 1c and 1d. The 3D position of the spherical cap can be stirred by scanning the tilt and azimuthal angles of the illumination[24], allowing us to computationally fill a certain volume of the 3D frequency space[2,22]. The expanded axial and lateral bandwidths of the 3D synthetic aperture result in the depth- and super-resolved imaging performance, respectively. In an extreme case, the half-pitch lateral and axial resolutions down to 90 and 150 nm have been realized, respectively[22].

**Experimental systems.** Our experimental implementations are schematically shown in Fig. 2. The VIS (10 ns duration, 532 nm wavelength) and MIR (1 μs duration, tunable wavenumber between 1,450 – 1645 $cm^{-1}$) lasers produce electrically synchronized optical pulse trains of 1 kHz repetition rate with a fixed time-delay of the former to the latter pulse (see Fig. 2a). The MIR light is intensity-modulated by a square wave such that the image sensor alternately captures the MIR ON and OFF frames at ~100 Hz. The MIR pulse fluence is ~10 pJ/$\mu m^2$ (~100 nJ over ~100 μm ×100 μm) but depends on the wavenumber. The VIS pulse fluence can be as low as ~0.1 pJ/$\mu m^2$ (~1 nJ over ~100 μm × 100 μm), which is 3 – 4 orders of magnitude lower than that used in, e.g., coherent Raman imaging[10].

Figure 2b shows our MV-DH system. DH is a wide-field interferometric technique to measure the 2D optical-phase-delay map[1,20,25]. The collimated laser beam illuminates the sample and its magnified complex-field image is replicated by the subsequent diffraction grating. The zeroth-order term is low-pass filtered to create a quasi-plane reference wave while the first-order term is transmitted unperturbed, such that these two terms create an interferogram on the image sensor. The illumination optical power is ~100 μW which is enough to use the full dynamic range of the image sensor that runs at 100 Hz. The mechanism of the diffraction limit of DH corresponds to the situation illustrated in Fig. 1c. The diffraction-limited half-pitch lateral resolution is ~440 nm as determined by the objective lens NA of 0.6. The temporal phase sensitivity of our DH system is dominated by the optical shot noise and ~10 mrad in standard deviation without averaging.

Figure 2c shows our MV-ODT system. ODT estimates the depth-resolved 3D RI map of the sample through the multi-angle tomographic measurements and computational reconstruction incorporating the diffraction effect[2,22]. The collimated VIS laser beam illuminates the sample with various incident angles stirred by the rotating wedge prism, and its magnified complex-field image is detected with the image sensor by means of Mach-Zehnder interferometry with the reference wave. We use the fixed illumination NA of 0.55 and scan 9 azimuthal angles with an increment of 36 degrees. The illumination optical power is ~1 μW which is enough to use the full dynamic range of the image sensor that runs at 60 Hz. The mechanism of the diffraction limit of ODT corresponds to the situation illustrated in Fig. 1d. The diffraction-limited half-pitch lateral and axial resolutions are ~190 nm and ~2.3 μm, respectively, as determined by the illumination and collection NAs of 0.55 and 0.85, respectively. The temporal RI sensitivity of our ODT system is dominated by the optical shot noise and ~2 × $10^{-5}$ in standard deviation without averaging.

**Basic performance of MV-QPI.** We first characterize the basic performance of our MV-QPI systems. The performance of our MV-DH system is summarized in our prior work[20]. The performance of our MV-ODT system is summarized in Fig. 3. We measure liquid oil sandwiched between two $CaF_2$ substrates of 500 μm thickness which is excited by the MIR beam with the focus diameter of ~30 μm. We average 500 MIR ON-OFF measurements for each photothermal tomogram. In Fig. 3a, we verify that the signal (i.e., the RI decrease) varies linearly against the MIR excitation energy. In Fig. 3b, we confirm exponential temporal decay of the signal with the decay constant of ~130 μs by varying the time-delay between the MIR and VIS pulses. Similar decay constants could be obtained with other experimental conditions as the thermal diffusivities of various liquids and polymers are in the order of $10^{-7}$ [$m^2$/s][26,27]. In Fig. 3c, we perform the MIR spectroscopy of the oil by scanning the MIR wavenumber. The obtained spectrum shows good agreement with the spectrum of the same oil obtained by a commercial Fourier-transform infrared spectrometer (FTIR).

**Comparison of the depth-resolving effects between MV-DH and MV-ODT.** In the context of MV-QPI, the depth-resolved quantitative RI imaging capability of ODT is critical for (1) decoupling the MIR photothermal effects in the out-of-focus aqueous layers and (2) quantifying the actual photothermal temperature change. We demonstrate these effects in Fig. 4 by comparing the photothermal images of fixed HEK293 cells immersed in $D_2O$-based phosphate-buffered saline (PBS) obtained with the DH and ODT reconstruction algorithms. To maintain the consistency in the FOV and spatial resolution, we apply the synthetic-aperture DH algorithm[23] on the same raw measurement dataset as those used in the ODT reconstruction. In short, the synthetic-aperture DH enhances the lateral resolution of DH by the same principle as the multi-angle ODT, but without mapping to the 3D frequency space. We average 2,500 MIR ON-OFF measurements, resulting in the acquisition time of 12.5 minutes. In this experiment, the MIR fluence is intentionally made non-uniform within the FOV to make a clearer comparison. In Fig. 4c, the cell's photothermal signals are contaminated by the non-uniform background originating from the water's photothermal phase change reflecting the MIR fluence. This background can be made more flattened by sectioning only the in-focus layer with the ODT algorithm, as shown by the blue regions of the cross-sectional profiles (see Figs. 4c and 4d). Also, some of the intracellular structures, as those indicated by the red regions in the cross-sectional profiles, are resolved with higher contrasts in the ODT reconstruction. Furthermore, ODT quantifies the photothermal RI change to be ~$10^{-5}$, from which the intracellular temperature rise can be estimated to be ~0.1 K assuming the water's thermo-optic coefficient of ~1.4 × $10^{-4}$ [1/K][28]. Note that the estimation of the temperature rise is generally not possible with MV-DH because the thickness and RI information are coupled in the obtained phase values.

**Live-cell, broadband MIR-fingerprint MV-QPI.** We demonstrate live-cell, broadband MIR-fingerprint imaging of a COS7 cell immersed in $H_2O$-based culture medium with the MV-DH system (see Fig. 5). We average 2,500 MIR ON-OFF frames for each spectral point, resulting in the acquisition time of 50 seconds. The water's MIR absorption effect is independently measured and subtracted. The QPI provided in Fig. 5a reveals the comprehensive morphology of the cell where the global cellular shape and various intracellular structures can be recognized such as nucleus, nucleoli and small cytoplasmic particles. We acquire MV-spectral images at 27 spectral points between 1,453 and 1,632 $cm^{-1}$. This spectral range resides in the MIR fingerprint region where $CH_2$ bending (1,450 – 1,500) and peptide bond's amide II (1,500 – 1,580) and amide I (1,580 – 1,700) bands show characteristic spectral signatures, which are recognized to be abundant in lipids and proteins, respectively[11]. Figure 5b shows the obtained MIR spectrum at different locations in the FOV. We can observe different spectral signatures across different intracellular structures; e.g., compared to the cytoplasm, the nucleolus shows a stronger signal of the amide II band centered at ~1,550 $cm^{-1}$. Indeed, the MV image of 1,548 $cm^{-1}$ clearly visualizes the signal localizations at the nucleoli which could represent the richness of proteins (see Fig. 5d).

Also, the small cytoplasmic localizations of 1,472 cm$^{-1}$ signal at the cellular boundary could represent the existence of lipid droplets (see Fig. 5c).

**Depth-resolved, broadband MIR-fingerprint MV-QPI.** We demonstrate depth-resolved, broadband MIR-fingerprint imaging of fixed HEK293 cells immersed in D$_2$O-based PBS with the MV-ODT system (see Fig. 6). We average 1,500 MIR ON-OFF measurements for each spectral point, resulting in the acquisition time of 7.5 minutes. We acquire MV-spectral tomograms at 19 spectral points between 1,502 and 1,632 cm$^{-1}$. The cross-sectional images of the sample's reconstructed 3D RI distribution at two different heights are shown in Figs. 6a and 6b, sectioning the podia (red arrow) and the nucleoli (red square) of the cells, respectively. The MIR spectrum obtained at the nucleolus resolves the spectroscopic signatures of the amide II band (see Fig. 6e) and its resonance at 1,563 cm$^{-1}$ is shown in Figs. 6c and 6d. The MV contrast is localized at the nucleoli in the top two cells, while it shows a characteristic cytoplasmic distribution surrounding the nucleus in the bottom cell which reminds us of an endoplasmic reticulum. Such intercellular variation, mainly of protein distribution, could represent different phases of the cell cycle. Finally, we can observe the depth-resolved localization of the MV signal and the RI contrast originating from the nucleolus in Fig. 6f. To the best of our knowledge, this is the first demonstration of the depth-resolved MV-QPI which is enabled by the implementation of the ODT scheme. The temperature rise inside the cells can be estimated to be ~0.1 K assuming the water's thermo-optic coefficient of ~1.4 × 10$^{-4}$ [1/K].

**Discussions**
The current MV-QPI systems have rooms for improvement. First, the spatial resolution of the photothermal images is limited by the heat diffusion that happens within the relatively long MIR pulse duration[29] (i.e., ~700 nm of the diffusion length in 1 μs). We estimate ~10 ns pulse duration confines the diffusion within the diffraction limit of the QPI systems. Second, a higher-energy MIR pulse laser is desired to increase the photothermal signal by an order of magnitude by creating the temperature rise of ~1 K. It is also desired to broaden the spectral tunability of the MIR light source so that various other biomolecules can be probed. In our experimental systems, these issues arise due to the use of the semiconductor MIR quantum cascade laser that offers low optical output power with a limited gain bandwidth. Third, a higher-full-well-capacity image sensor[30] can be used to enhance the shot-noise-limited sensitivities of our MV-QPI systems by at most two orders of magnitude. Fourth, any other QPI methods can be implemented to harness their versatility[5], robustness[31-33] and imaging speed[34-37] while higher-NA objective lenses can be used to enhance the lateral and axial spatial resolutions[22].

With the above-mentioned improvements, MV-QPI method has the potential to achieve high-sensitive, low-photodamage and super-resolution MVI. In total, nearly three orders of magnitude enhancement in the MV detection sensitivity can be expected. In this case, the VIS fluence at the sample plane increases by ~3 orders of magnitude and becomes comparable to that used in coherent Raman imaging[10]. Our method can still reduce photodamage associated with nonlinear electronic transitions of biomolecules[21] since the duration of the VIS pulse can be ~10 ns which is significantly longer than the picosecond or femtosecond pulses used in coherent Raman imaging. Also, with the reduced MIR pulse duration, the spatial resolution of the photothermal images can, in principle, reach diffraction limit of the synthetic-aperture QPI system, which surpasses the diffraction limit posed in other MVI techniques[6-18,20] by the NA of a single objective lens.

## Methods

**Light sources.** The VIS light source is the second harmonic generation (SHG) of 10 ns, 1 kHz, 1,064 nm pulsed laser beam emitted from the Q-switch laser NL204-1K (Ekspla) produced with the nonlinear crystal LBO-503 (Eksma Optics). The spatial mode of the SHG beam is cleaned by the single-mode optical fiber P3-405B-FC-5 (Thorlabs). The MIR light source is DO418 (Hedgehog, Daylight Solutions) providing access to 1,450 - 1,640 $cm^{-1}$ region with the pulse duration of ~1 μs.

**Measurement of the MIR pulse-energy spectrum.** The intensity spectrum of the MIR light source is generally not uniform; hence, the raw photothermal (i.e., MIR ON-OFF) contrast needs to be normalized by the corresponding MIR pulse energy in order to obtain the sample-specific spectroscopic information. We use a cadmium-mercury telluride detector PVM-10.6-1x1-TO8 (VIGO System) to measure the waveform of the MIR pulse at each wavenumber with an oscilloscope TDS2024C (Tektronix), and use the area of the obtained waveform as a measure of the pulse energy. This measurement is performed independent of the image acquisition.

**MV-DH system.** The image sensor is operated at 100 Hz with the exposure time of 9 ms. The MIR ON-OFF modulation rate is 50 Hz. The VIS illumination power at the sample plane is ~100 μW which is enough to use the full dynamic range of the image sensor. The MIR pulse energy at the sample plane is ~100 nJ on average but depends on the wavenumber (e.g., ~200 nJ at 1,548 $cm^{-1}$).

**MV-ODT system.** The image sensor is operated at 60 Hz with the exposure time of 15 ms. The MIR ON-OFF modulation rate is 30 Hz. The VIS illumination power at the sample plane is ~1 μW which is enough to use the full dynamic range of the image sensor. The MIR pulse energy at the sample plane is ~100 nJ on average but depends on the wavenumber (e.g., ~200 nJ at 1,548 $cm^{-1}$).

**Materials used to characterize the basic performance of the MV-ODT system.** Series A 1.54000 (Cargille) is used as the liquid oil sample. FT/IR-6800, ATR PRO ONE and PKS-D1F (JASCO) are used to obtain the reference MIR absorption spectrum of the oil.

**Biological samples.** The COS7 cells (Riken) are cultured in Dulbecco's Modified Eagle's Medium (DMEM) with 10% fetal bovine serum supplemented with penicillin–streptomycin, L-glutamine, sodium pyruvate and nonessential amino acids at 37 °C in 5% $CO_2$. For the live-cell imaging, the cells are cultured in 35-mm glass-bottomed dishes (AGC Techno Glass) and the medium is replaced by phenol red-free culture medium containing HEPES buffer (2 mL) before imaging. All solutions are from Thermo Fisher Scientific. The HEK293 cells are fixed with 4% paraformaldehyde at room temperature for 5 minutes and immersed in $D_2O$-based PBS before observation. All the samples are sandwiched between two $CaF_2$ substrates of 500 μm thickness before observation.

## Data availability
The data provided in the manuscript are available from T.I. upon request.

## Acknowledgements
We thank Makoto Kuwata-Gonokami and Junji Yumoto for letting us use their equipment. This work was financially supported by JST PRESTO (JPMJPR17G2) and JSPS KAKENHI (17H04852, 17K19071).

## Author contributions
T.I. conceived the concept. M.T. proposed MV-ODT, designed and constructed the systems, designed and performed the experiments and analyzed the experimental data. K.T. contributed in design of the systems and the experiments, interpretation of the obtained data and daily discussion regarding the MV-QPI method. H.S. wrote the automated data acquisition software program and the graphical user interface. T.H., M.K. and K.O.

prepared the COS7 cell. Y.N. prepared the HEK293 cells and provided biological perspectives for the data interpretation. R.H. advised on the computational reconstruction framework of ODT. T.I. supervised the entire work. M.T. and T.I. wrote the manuscript with contributions from all the other authors.

**Competing interests**
Authors declare no competing interest.

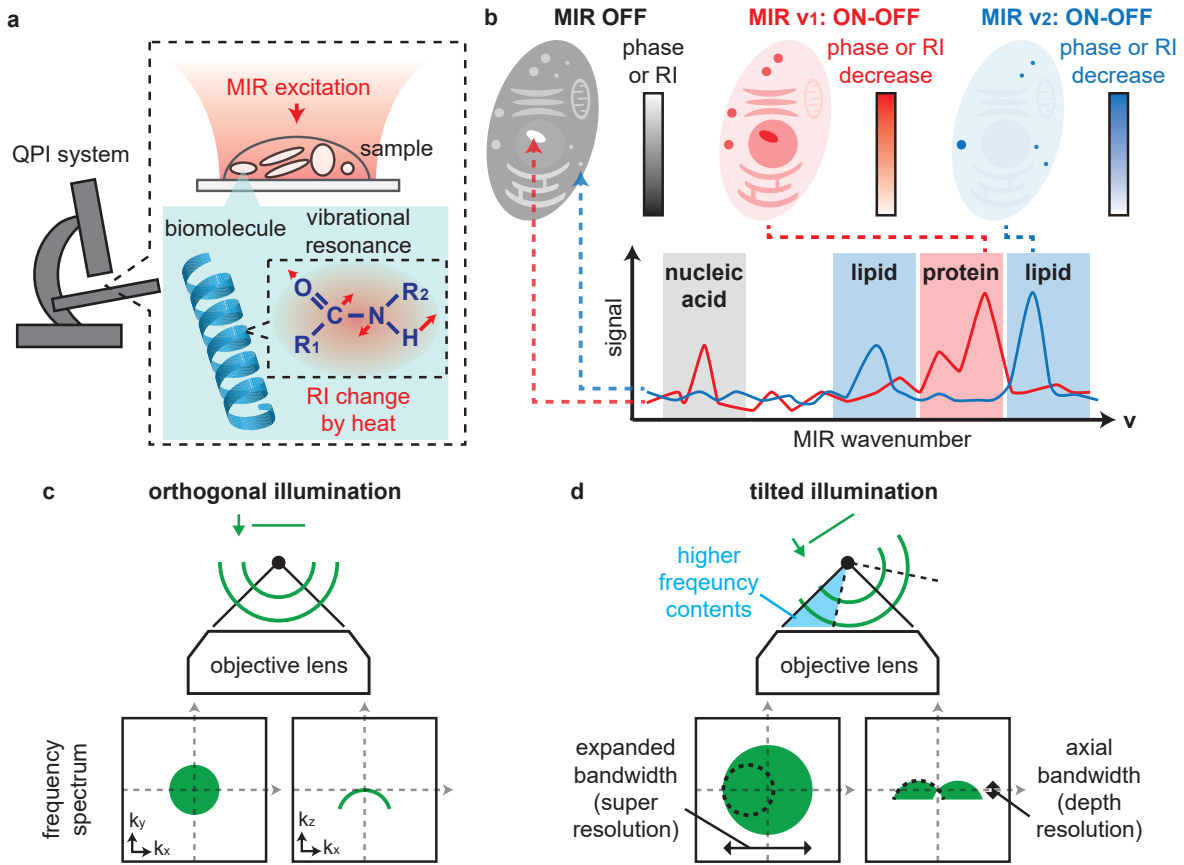

**Fig. 1 | Concept of MV-QPI. a**, Principle of the MV-contrast acquisition in the QPI framework. The MIR light of a certain wavenumber is irradiated to the entire volume of the sample, where the resonant biomolecules are selectively excited to their fundamental vibrational states. The vibrational energy is eventually transformed into heat that diffuses into the surrounding medium. The resulting photothermal RI decrease is detected by the QPI system with the spatial resolution of the VIS probe light. **b**, Cross-correlative analysis enabled by MV-QPI. The phase or RI image obtained at the MIR OFF state reveals the quantitative and comprehensive morphology of the sample containing rich information about cellular shapes and intracellular-organelle distributions. Scanning of the MIR wavenumber visualizes contrasts of various MV resonances at each spatial point of the FOV, which can be decomposed into individual biomolecular constituents through chemometric analysis. **c**, Mechanism of the diffraction limit in the standard 2D QPI. The object is illuminated with the orthogonal plane wave and only a limited range of spatial frequency information of the diffracted field can be collected with the NA of the objective lens. **d**, Mechanism of the depth- and super-resolution in the synthetic-aperture QPI. The object is illuminated with the tilted plane wave such that higher-frequency contents can be collected that used to be outside the NA of the objective lens in **c**. Scanning the tilt and azimuthal angles of the illumination allows us to computationally synthesize the 3D frequency aperture. The depth- and super-resolved imaging performance can be achieved with the expanded lateral and axial bandwidths of the 3D synthetic aperture, respectively. The black dotted curves in the frequency spectrum indicate the Ewald's spherical cap that determines the 3D coverage of the objective lens' NA under a certain angle of illumination.

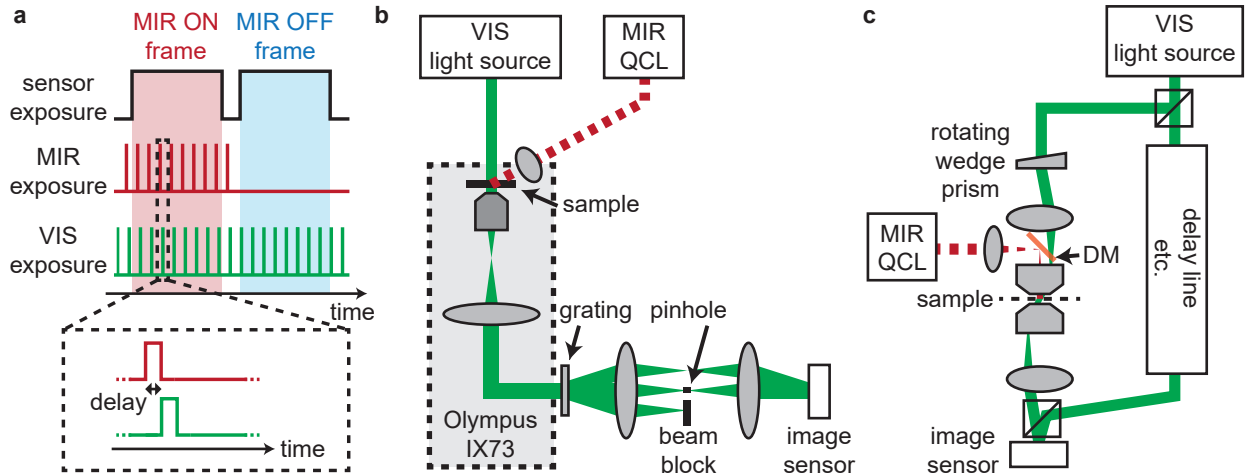

**Fig. 2 | Experimental implementations. a**, Temporal synchronization. The VIS and MIR lasers are electrically controlled to synchronize their pulse repetitions (~1 kHz) and relative time-delay. The MIR beam is intensity-modulated to be in-phase with the half-harmonic of the image sensor's frame rate (~100 Hz). **b**, MV-DH system. The DH microscope is built based on a commercial microscope housing IX73 (Olympus). The collimated VIS laser beam is used as the plane-wave probe illumination and the magnified image of the sample is formed next to the output port of IX73. The subsequent 4f system is used to perform the common-path off-axis interferometry. The MIR laser beam is loosely focused onto the sample with a $CaF_2$ lens. QCL: quantum cascade laser. **c**, MV-ODT system. The collimated VIS laser beam is split into two paths to create the Mach-Zehnder off-axis interferometer. The deflection angle of the probe beam created by the wedge prism is magnified and relayed into the sample plane by the subsequent tube lens and the illumination objective lens. The resulting tilted plane-wave illumination is then collected by another objective lens and the subsequent tube lens forms the sample's magnified image on the image sensor. The MIR and VIS beams are combined by the dichroic mirror (DM).

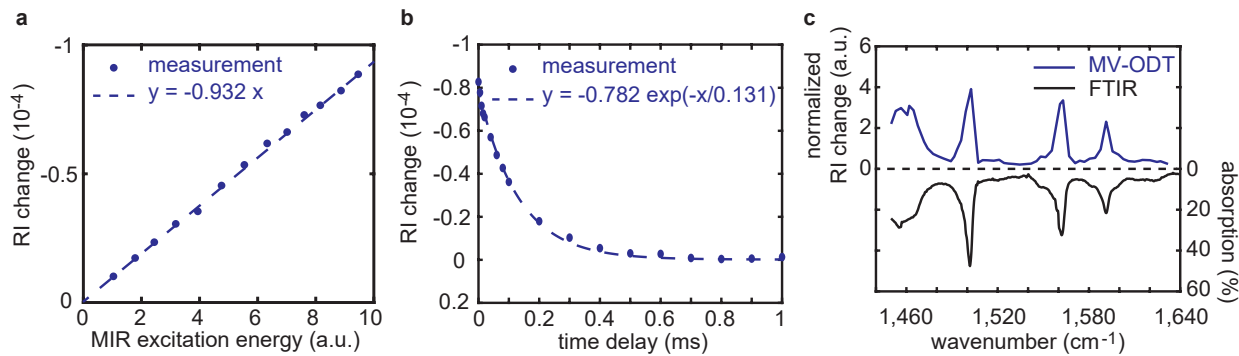

**Fig. 3 | Basic performance of the MV-ODT system.** The liquid oil sandwiched between two $CaF_2$ substrates is used as the sample which is excited by the MIR beam with the focus diameter of ~30 μm. **a**, Linearity of the photothermal RI change with respect to the MIR excitation pulse energy. **b**, Exponential temporal decay of the photothermal RI change with the decay constant of ~130 μs. **c**, MIR spectrum of the liquid oil obtained by the MV-ODT system, showing good agreement with the FTIR reference spectrum. Each measurement point shown in **a** – **c** represents one voxel of the FOV.

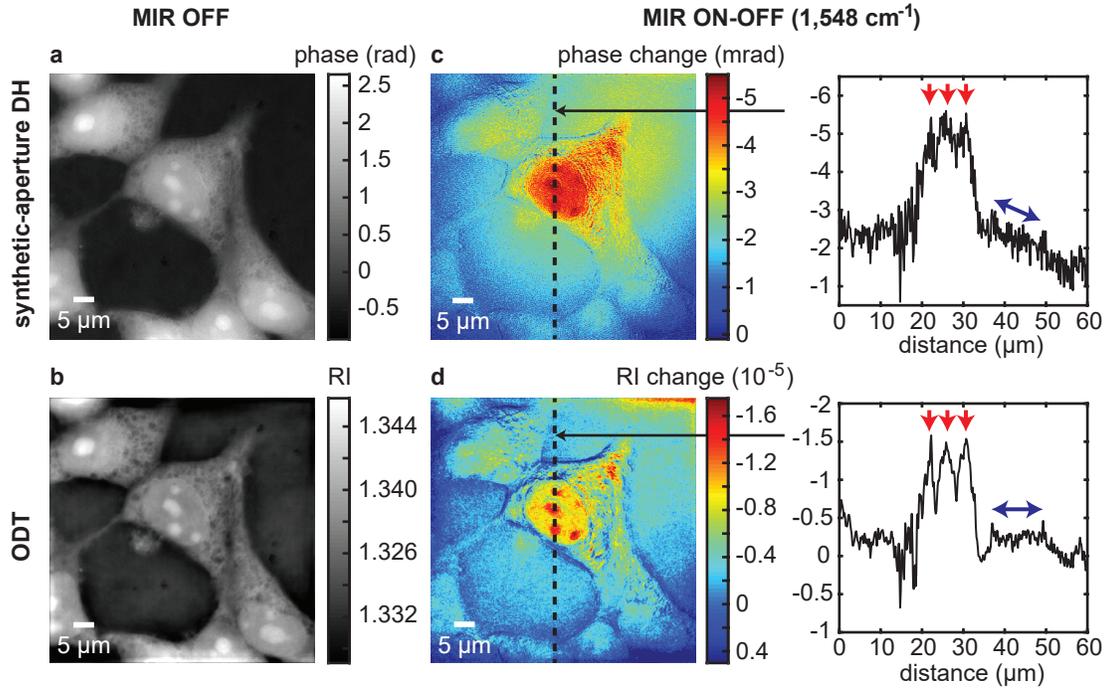

**Fig. 4 | Comparison of the depth-resolving effects between MV-DH and MV-ODT. a**, **b**, Raw phase and RI images of the fixed HEK293 cells in $D_2O$-based PBS at the MIR OFF state, respectively. **b** sections one particular height of the reconstructed 3D RI tomogram. **c**, **d**, Photothermal contrasts of the same FOVs as those shown in **a** and **b**, respectively, obtained with the MIR wavenumber tuned to 1,548 cm$^{-1}$. In **c**, the cellular structures are contaminated by the photothermal signals originating from the out-of-focus aqueous layers. In **d**, the depth-resolution provided by ODT results in the higher contrasts of the cellular structures (indicated by the red arrows) as well as the more uniform and flattened background distribution originating from the MIR absorption of the in-focus water layer (indicated by the blue arrows). The depth-resolved quantification of the RI values also allows for accurate estimation of the photothermal temperature rise inside the cells (~0.1 K) using the thermo-optic coefficient of water (~1.4 ×10$^{-4}$ [1/K]). In this experiment, the MIR fluence is intentionally made non-uniform within the FOV to make a clearer comparison.

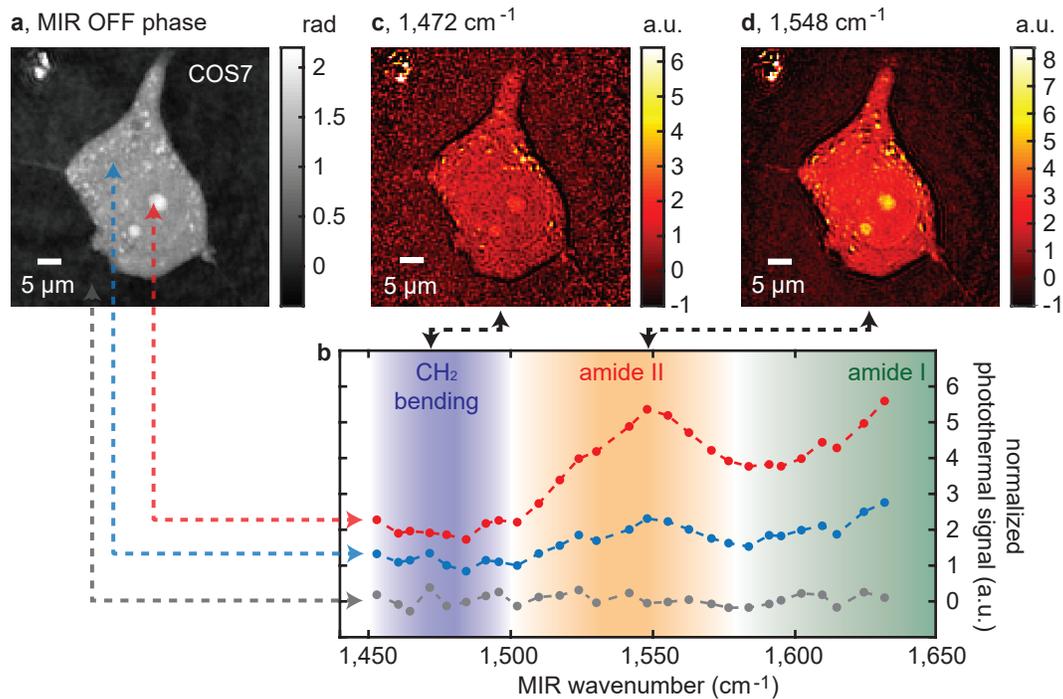

**Fig. 5 | Live-cell, broadband MIR-fingerprint MV-DH microscopy. a**, Raw phase image of the live COS7 cell in H$_2$O-based culture medium at the MIR OFF state. **b**, MIR spectrum of the nucleolus (orange), cytoplasm (blue) and empty area (gray) indicated by the arrows of the respective colors in **a**. The scanned MIR wavenumber region resides in the MIR fingerprint region where spectroscopic signatures of CH$_2$ bending and peptide bond's amide I and II bands can be found, which are abundant in lipids and proteins, respectively. Compared to the cytoplasm, the nucleolus shows the stronger signal of the broad absorption centered at ~1,550 cm$^{-1}$ which coincides with the amide II band. Each spectral points represents the spatial average of 3 × 3 diffraction-limited pixels (1.3 μm × 1.3 μm). **c**, **d**, MV images of the cell resonant to 1,472 and 1,548 cm$^{-1}$, respectively, after normalization. In **c**, the small cytoplasmic localizations of the MV contrast at the cellular boundary could represent the existence of lipid droplets. In **d**, the MV contrast is selectively strong on the nucleoli which could represent the richness of proteins.

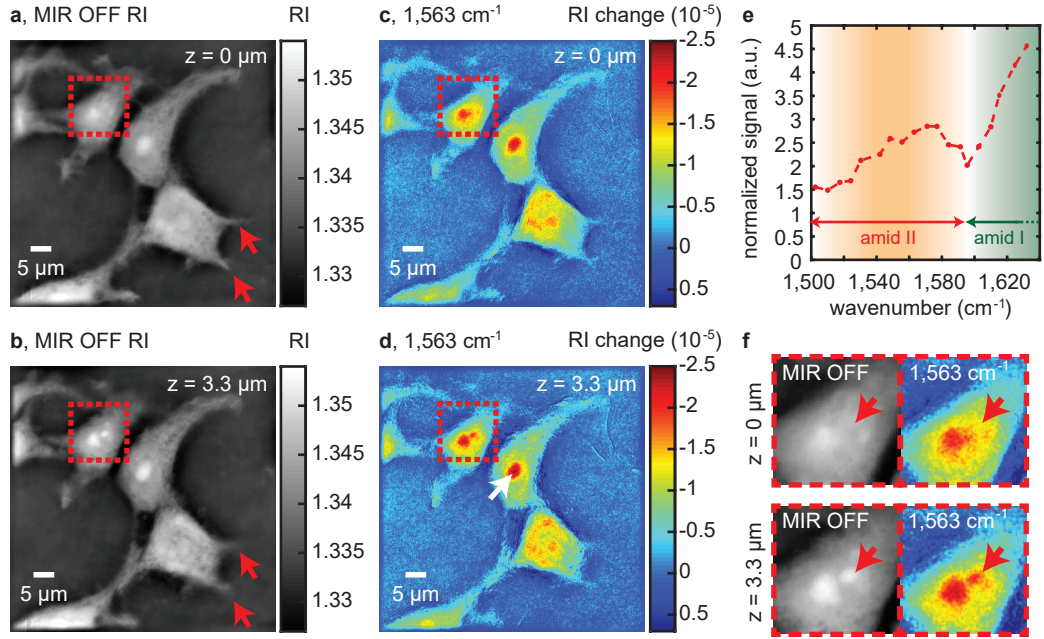

**Fig. 6 | Depth-resolved, broadband MIR-fingerprint MV-ODT microscopy. a**, **b**, Cross-sectional images in two different axial planes of the reconstructed RI tomogram of the fixed HEK293 cells in $D_2O$-based PBS at the MIR OFF state. **a** sections the podia of the cell (red arrow) while **b** the nucleoli (red square). **c**, **d**, MV contrasts of the same FOVs as those shown in **a** and **b**, respectively, resonant to 1,563 cm$^{-1}$. The photothermal temperature rise inside the cells can be estimated to be ~0.1 K using the thermo-optic coefficient of water. **e**, MIR spectrum at one voxel of the FOV in the nucleolus indicated by the white arrow in **d**, resolving the characteristic signature of the amide II band. **f**, Enlargement of the red-square regions in **a** – **d**. At z = 0 μm, the nucleolus indicated by the red arrows are not visible in the RI or the photothermal contrast. At z = 3.3 μm, the nucleolus appears in the RI contrast which also gives the signal in the photothermal contrast, demonstrating the depth-resolving capability.